# Field-Induced Spin-Density-Wave Phases in Quasi-One-Dimensional Conductors: Theory versus Experiments

A. G. Lebed

*L. D. Landau Institute for Theoretical Physics RAS, 2 Kosygina Street, 117334 Moscow, Russia*

We show that the "quantized nesting" model misses important features of the magnetic-field-induced spin-density-wave (FISDW) phase diagram. Among them are the following: (1) the FISDW wave vector is not strictly quantized; (2) in some compounds, the FISDW diagram consists of two regions: (a) At low temperatures, there are jumps of the wave vector (i.e., the first order transitions between FISDW phases); (b) at high temperatures the jumps and the first order transitions disappear, but the wave vector is still a nontrivial function of a magnetic field. These are in agreement with the experiments on $(TMTSF)_2PF_6$.

Magnetic-field-induced spin-density-wave (FISDW) phenomenon has been intensively studied both experimentally and theoretically [1–3] since its discovery in quasi-one-dimensional (Q1D) organic compounds $(TMTSF)_2X$ (where TMTSF denotes tetramethyltetraselenafulvalene, $X = PF_6$, $ClO_4$, $AsF_6$, etc.) [1–9] and $(DMET\text{-}TSeF)_2X$ ($X = AuI_2$ and $AuCl_2$) [10,11]. The best studied compounds, $(TMTSF)_2X$ ($X = PF_6$, $ClO_4$), are characterized by a simple Q1D electron spectrum under pressure [1–3,8]:

$$\epsilon^{\pm}(\mathbf{p}) = \pm v_F(p_x \mp p_F) - 2t_b \cos(p_y b^*) - 2t_b' \cos(2p_y b^*) - 2t_c \cos(p_z c^*), \quad (1)$$

where + (−) stands for the right (left) sheet of the Fermi surface (FS); $v_F$ and $p_F$ are the Fermi velocity and Fermi momentum; $t_b \simeq 200$ K, $t_b' \simeq 10$ K, and $t_c \simeq 5$ K are the overlapping integrals between electron wave functions [1–3]; $\hbar \equiv 1$. It is important that at $t_b' = 0$ the left sheet of the FS (1) coincides with the right one if it is shifted by a nesting vector:

$$\mathbf{Q}_0 = (2p_F, \pi/b^*, \pi/c^*). \quad (2)$$

Nesting properties (2) of the Q1D FS (1) [i.e., small enough value of the "antinesting term," $t_b'$, in Eq. (1)] and the Peierls instability [1–3] were suggested [1–3,12] to be responsible for a SDW ground state in $(TMTSF)_2PF_6$ at $T \leq 12$ K, $H = 0$, and ambient pressure. The first theories [13,14] of the FISDW phenomenon demonstrated that an effective space dimensionality of the Q1D electrons (1) reduces in a magnetic field [13,14]. Therefore, the Peierls instability [1] results in the appearance of the FISDW phases even in such compounds where the antinesting term, $t_b'$, destroys the SDW state at $H = 0$ [13,14].

An improved theory, a so-called "quantized nesting" (QN) model, was elaborated [1–3,15–24] on the basis of the results [13,14]. A keystone statement of the QN model is that a longitudinal projection of the FISDW wave vector is quantized:

$$\Delta_{Q_N}(\mathbf{r}) = \Delta_{Q_N} \exp(i\mathbf{Q}_N\mathbf{r}), \quad \mathbf{Q}_N = [2p_F + 2N\omega_c(H)/v_F, \pi/b^*, \pi/c^*], \quad (3)$$

where $N$ is an integer, $\omega_c(H) = \frac{eHv_F b^*}{c}$ is a frequency of an electron motion along open FS (1) in a magnetic field,

$$\mathbf{H} = (0, 0, H), \qquad \mathbf{A} = \left(0, \frac{e}{c}Hx, 0\right), \quad (4)$$

applied perpendicular to the conducting chains of a Q1D compound (1), where $e$ is the electron charge, $c$ is the velocity of light.

According to the QN model, each "quantized" FISDW phase (3) is characterized by its own metal-FISDW transition temperature, $T_{\mathrm{FISDW}}(N,H)$ [1–3,15,19,20], and by its own free energy, $F_{\mathrm{FISDW}}(N,H,T)$ [1,2,16–18], which depend on the integer $N$ [see Eq. (3)]. Thus, to determine the FISDW phase diagram in a framework of the QN model, it is necessary to find the largest value of $T_{\mathrm{FISDW}}(N,H)$ at fixed $H$ and the lowest value of $F_{\mathrm{FISDW}}(N,H,T)$ at fixed $H$ and $T$. As a result, a cascade of the first order phase transitions between the FISDW phases characterized by different longitudinal wave vector (3) was theoretically suggested [1–3,15–24] [see Fig. 1(a)]. An important consequence of the "quantization rule" (3), a so-called "three-dimensional quantum Hall effect" (3D QHE), is experimentally observed in $(TMTSF)_2X$ [4–9] and $(DMET\text{-}TSeF)_2X$ [11] compounds and is theoretically explained in Refs. [21–24].

We stress that the calculations [1–3,15–24] which established the QN model [i.e., Eq. (3)] were performed only for very low metal-FISDW transition temperatures, $[\pi T_{\mathrm{FISDW}}(N,H)] \ll \omega_c(H)$. Our main message is that the QN model has to be considered only as a limiting case (where $[\pi T_{\mathrm{FISDW}}(N,H)]/\omega_c(H) \to 0$) of our more general analysis of the equation defining $T_{\mathrm{FISDW}}(N,H)$ derived in Ref. [15] (see also Ref. [25]). In other words, we claim that numerous applications of the QN model [15–24] to real FISDW phase diagrams in Q1D compounds have to be theoretically improved.

We demonstrate that, due to an electron-hole asymmetry of an electron spectrum in the FISDW phase with $N \neq 0$ [21], an account of a finite transition temperature, $T_{\mathrm{FISDW}}(N,H) \neq 0$, changes the main qualitative consequences of the QN model. In contrast to the QN model, we show the following: (1) The longitudinal wave vector of the FISDW phases is not strictly quantized [i.e., $N$ is not an integer in Eq. (3) unless $N = 0$]. In particular, this means that all existing calculations of the 3D QHE [20–24] have to be improved since they are based on a hypothesis about strict quantization of the wave vector (3) and treat the parameter $N$ as a topological number. (2) For small enough values of the ratio $h = \omega_c(H)/[\pi T_{\mathrm{FISDW}}(N,H)] < h_c$, the FISDW phase diagram consists of two regions: (a) a low-temperature region ("quantum FISDW") where there exist discontinuous (but noninteger) jumps of the FISDW wave vector (i.e., the first order transitions between different FISDW phases) [see Fig. 1(b)]; (b) a high temperature region ("quasiclassical FISDW") where the jumps and the first order transitions disappear, but the FISDW wave vector is still a nontrivial oscillating function of a magnetic field [see Fig. 1(b)]. (3) For large enough values of the ratio $h = \omega_c(H)/[\pi T_{\mathrm{FISDW}}(N,H)] > h_c$, the FISDW phase diagram consists of a cascade of the first order phase transitions between the different FISDW phases. They are characterized by discontinuous (but noninteger) jumps of the FISDW wave vector [see Fig. 1(c)]. Note that a special case of the $(N \simeq 1) \rightarrow (N = 0)$ phase transition where there is an electron-hole symmetry in the $N = 0$ phase [1,21] was studied by us earlier [25].

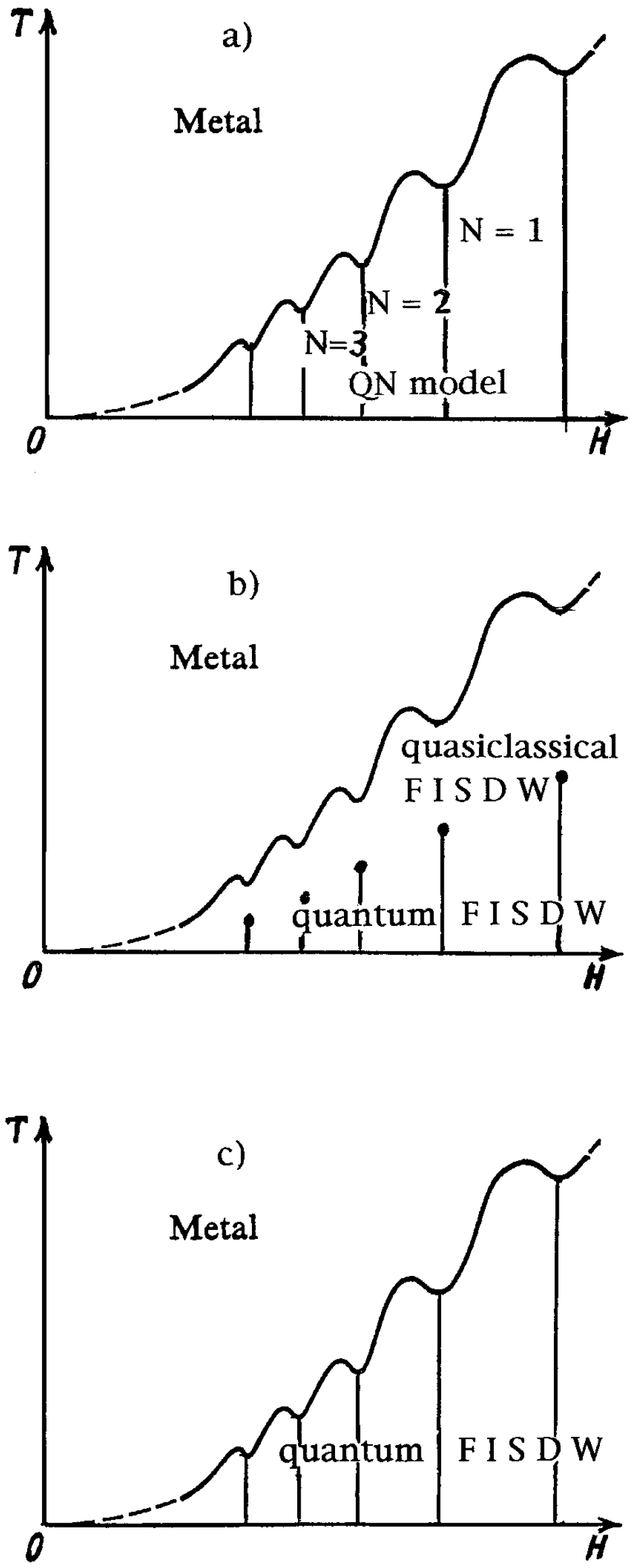

FIG. 1. (a) FISDW phase diagram derived from the QN model [13–24] consists of a cascade of the first order FISDW transitions where the integer $N$ [defining the FISDW phase wave vector (3)] exhibits integer jumps on the phase transition lines. (b) FISDW phase diagram suggested in the Letter for $h < h_c$ consists of two regions: the "quantum FISDW" [where there are the first order transitions between the FISDW phases corresponding to noninteger jumps of the parameter $N$ in Eq. (3)] and the "quasiclassical FISDW" (where the first order transitions disappear). (c) FISDW phase diagram suggested in the Letter for $h > h_c$ corresponds to noninteger jumps of the parameter $N$ in Eq. (3) on the first order FISDW phase transition lines (see the text).

In the Letter we consider a typical for the FISDW phase diagram phase transition $(N \simeq 2) \rightarrow (N \simeq 1)$ where there are no electron-hole symmetries in both phases, $N \simeq 2$ and $N \simeq 1$, to make conclusions about the whole FISDW phase diagram. As usual [1–3,12–24], we suggest that the antinesting term, $t_b'$, in the electron spectrum (1) destroys the SDW phase at $H = 0$. Let us study the FISDW phases which appear at high enough magnetic fields by means of the equation defining $T_{\mathrm{FISDW}}(N,H)$ for the Q1D electron spectrum (1) [1–3,15]:

$$\frac{1}{g} = \int_d^\infty \frac{2\pi T_{\mathrm{FISDW}}(N,H)\, dZ}{v_F \sinh[\frac{2\pi T_{\mathrm{FISDW}}(N,H) Z}{v_F}]} \times J_0\left(\frac{4t_b'}{\omega_c(H)} \sin\left[\frac{\omega_c(H) Z}{v_F}\right]\right) \cos(k_x Z), \quad (5)$$

where $J_0(\ldots)$ is the Bessel function, $g$ is an effective interaction constant, and $d$ is a cutoff distance. Unlike the previous calculations [1–3,15–24] [where only the "quantized values" of the wave vector, $k_x = 2N\omega_c(H)/v_F$, are considered], below we consider a continuous variable $N$ and find a maximum of the FISDW transition temperature, $T_{\mathrm{FISDW}}(N,H)$, at given $H$ and the parameter $h = \omega_c(H)/[\pi T_{\mathrm{FISDW}}(N,H)]$.

Using the following mathematical equalities [26]:

$$J_0(2x \sin\alpha) = \sum_{l=-\infty}^{+\infty} J_l^2(x) \exp(2il\alpha), \quad (6)$$

$$\frac{1}{\sinh(y)} = 2 \sum_{m=0}^{+\infty} \exp[-y(2m+1)], \quad (7)$$

it is possible to rewrite Eq. (5) in the form more suitable for numerical analysis [27]:

$$\ln\left[\frac{T_0}{T_{\mathrm{FISDW}}(N,H)}\right] = 2 \sum_{l=-\infty}^{+\infty} J_l^2(\lambda') \sum_{M=0}^{+\infty} \frac{(H_N^*)^2}{(2M+1)[(2M+1)^2 + (H_N^*)^2]}, \quad (8)$$

$$H_N^* = \frac{\omega_c(H)}{\pi T_{\mathrm{FISDW}}(N,H)}\,(l - N)\,, \tag{9}$$

where $\lambda' = 2t_b'/\omega_c(H)$, $T_0$ is a SDW phase transition temperature at $H = t_b' = 0$. [Note that Eqs. (8) and (9) are mathematically equivalent to Eq. (5) [1–3,15] and, thus, they define a metal-FISDW phases transition line. Another point is that the ratio $h = \omega_c(H)/[\pi T_{\mathrm{FISDW}}(N,H)]$ is pressure dependent [8] and, thus, can be considered as an independent parameter.]

Let us solve Eqs. (8) and (9) numerically and find some critical value of the parameter $h$, $h_c^{1,2}$. We define $h_c^{1,2}$ as follows: at $h > h_c^{1,2}$ there exists a tricritical point corresponding to an intersection of the metal-FISDW second order phase transition line with the $(N \simeq 2) \to (N \simeq 1)$ first order phase transition line [see Fig. 1(c)], whereas at $h < h_c^{1,2}$ this tricritical point disappears [see Fig. 1(b)]. To estimate the value of $h_c^{1,2}$, one has to minimize the right-hand side of Eq. (8) at different fixed values of the parameter $h$ and to find the function $N = N(\lambda', h)$. As it follows from Eqs. (8) and (9), this procedure gives the maximum values of the metal-FISDW transition temperature, $T_{\mathrm{FISDW}}(\lambda', h)$. Then, one has to investigate the functions $N = N(\lambda', h)$ in order to find the critical value, $h_c^{1,2}$, where the tricritical point disappears.

The results of our numerical calculations are shown in Fig. 2 and the corresponding FISDW phase diagrams are presented in Figs. 1(b) and 1(c). We find that the critical value of the parameter $h$ is $h_c^{1,2} \simeq 1$ for the $(N \simeq 2) \to (N \simeq 1)$ FISDW phase transition [see Fig. 2(b)]. Indeed, as it is seen from Fig. 2(a), at $h = 1.2$ there is a jump of the variable $N$ and, thus, a jump of the FISDW wave vector (3) corresponding to the first order $(N \simeq 2) \to (N \simeq 1)$ FISDW phase transition (we call such FISDW phases the "quantum FISDW" ones). On the other hand, at $h = 1$ and 0.8 the above-mentioned jump does not exist anymore and, thus, the first order $(N \simeq 2) \to (N \simeq 1)$ FISDW phase transition disappears (we call such FISDW phases the "quasiclassical FISDW" ones) [see Figs. 2(b) and 2(c)]. Nevertheless, we stress that even at $h = 1$ and 0.8 the FISDW wave vector (3) still exhibits a very peculiar behavior. A detailed study of the quasiclassical FISDW states, including a possibility of the existence of some phase transitions or crossovers between them, is not a subject of our Letter, and the corresponding results will be published elsewhere [28]. We also stress that there is another variant [1,19] of the QN model which accounts for the possible changes in a perpendicular component of the wave vector, $Q_y(H)$, with changing magnetic field. Its main predictions are the same as the predictions of the QN model considered in the Letter: the existence of a cascade of the FISDW first order phase transitions and the 3D QHE due to a similar quantization rule for a longitudinal projection of the FISDW wave vector:

$$\mathbf{Q}_N = [2p_F + N\omega_c(H)/v_F, Q_y(H), \pi/c^*]. \tag{10}$$

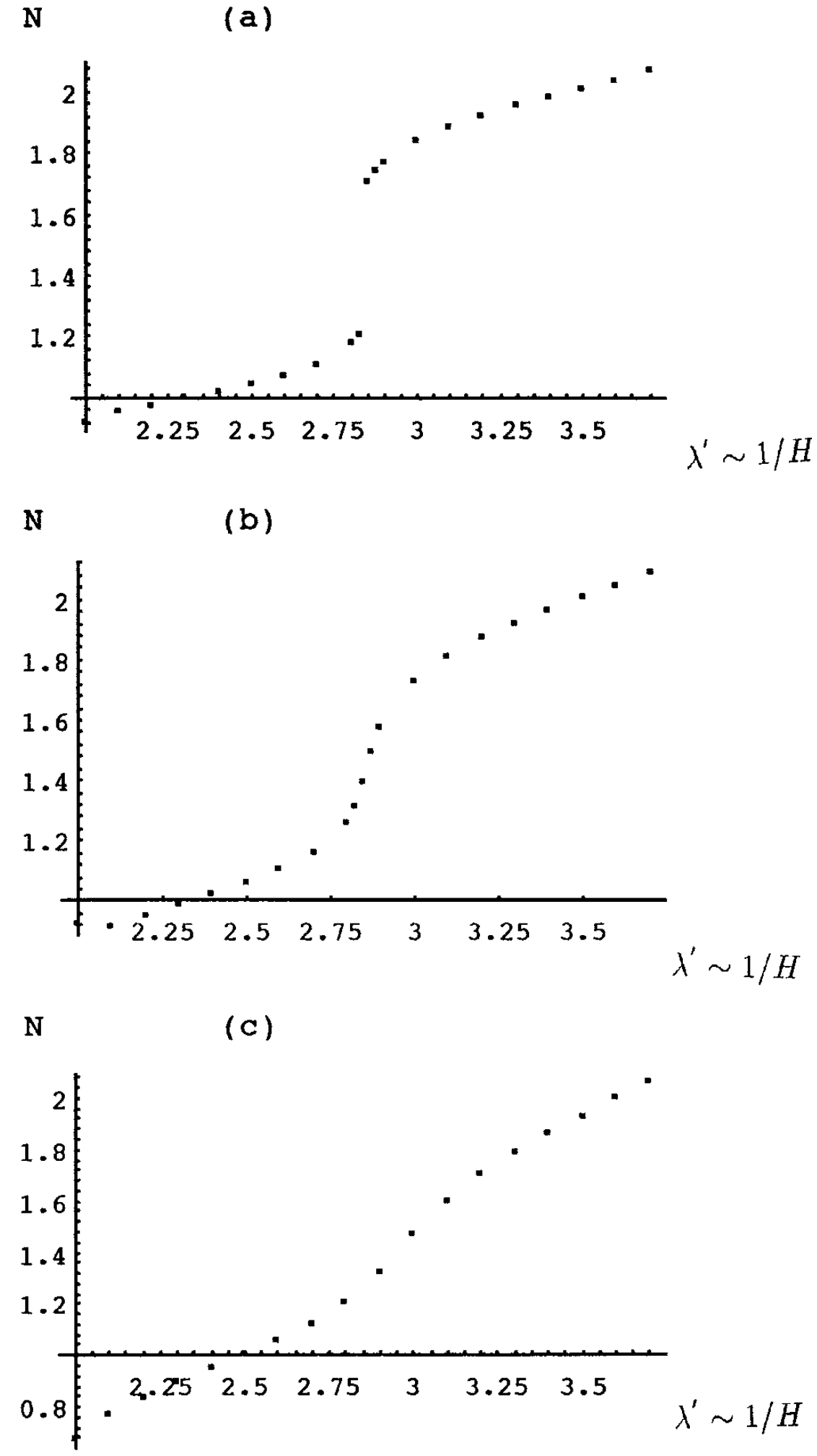

FIG. 2. (a) As shown in the Letter, $N$ [see Eq. (3)] exhibits a noninteger jump at $h > h_c^{1,2}$. (b) At $h = h_c^{1,2} \simeq 1$ the jump of $N$ first disappears. (c) Although at $h < h_c^{1,2}$ the jump does not exist anymore; nevertheless, the value of $N$ is still a nontrivial function of $1/H$.

We point out that the quantization rules (10) and (3) have been theoretically proved only in the limiting case where $[\pi T_{\mathrm{FISDW}}(N,H)]/\omega_c(H) \to 0$. Although a detailed study of the model [19] is not a subject of the Letter, our preliminary results [28] show that the qualitative statements of the Letter are retained in the model [19]. For example, at $[\pi T_{\mathrm{FISDW}}(N,H)/\omega_c(H)] = 1$, the minimum of the free energy in model [19] corresponds to a noninteger $N = 0.53$ [28]. Thus, we conclude that both quantization rules (3) and (10) are not strictly valid at finite transition temperatures, $\pi T_{\mathrm{FISDW}}(N,H)/\omega_c(H) \neq 0$.

Below, we compare our results with the classical [6,7] and recent [29] experiments on a compound $(TMTSF)_2PF_6$. If one takes the experimental estimation of $v_F \simeq 10^7$ cm/sec [30], one obtains the following estimation of the parameter $h$: $h_{1,2}^* = \omega_c(H)/[\pi T_{\mathrm{FISDW}}(H)] \leq 1$. Thus, one may expect that the FISDW phase diagrams in $(TMTSF)_2PF_6$ [6,7,29] consist of two regions: the quantum FISDW and quasiclassical FISDW. Indeed, as mentioned in Ref. [6], hysteresis characterizing the first order FISDW phase transitions disappears at temperatures lower than the metal-FISDW transition temperature, $T_{\rm FISDW}(H,N)$. In a very recent paper [29], this problem is studied in detail, and the line where the hysteresis disappears is shown to subdivide the FISDW diagram into two different areas. In Ref. [7], it is shown that the peaks of a resistivity determining the FISDW phase transitions disappear well below $T_{\rm FISDW}(H,N)$. Therefore, we summarize that the experiments [6,7,29] are in agreement with the analysis suggested in the Letter (see Figs. 1 and 2) and are in a qualitative disagreement with the results of the QN model [15–24] [see Fig. 1(a)].

In conclusion, we point out that, in our opinion, the temperature dependences of the 3D QHE plateaus measured in Ref. [7] and discussed in Ref. [31] have peculiarities on some lines inside the FISDW phase diagram. We speculate that the 3D QHE exists only in the quantum FISDW phases and disappears in the quasiclassical FISDW states, although this problem has to be carefully studied both theoretically and experimentally. We stress that the improvement of the existing theoretical descriptions of the FISDW phases suggested in the Letter gives rise to a discussion about the physical nature of the 3D QHE since its current descriptions [20–24] are based on a hypothesis that the parameter $N$ is an integer topological number which is not the case, as shown in the Letter. We also expect that such physical properties as a nonlinear FISDW conductivity and "magic angle" phenomenon are different in the quantum FISDW and quasiclassical FISDW phases (see Fig. 1) and suggest to prove these experimentally.

I am thankful to N. N. Bagmet, E. V. Brusse, P. M. Chaikin, T. Ishiguro, M. V. Kartsovnik, A. V. Kornilov, M. J. Naughton, V. M. Pudalov, and K. Yamaji for fruitful discussions.